\documentclass[fleqn,usenatbib]{mnras}

\usepackage{newtxtext,newtxmath}

\usepackage[T1]{fontenc}

\DeclareRobustCommand{\VAN}[3]{#2}
\let\VANthebibliography\thebibliography
\def\thebibliography{\DeclareRobustCommand{\VAN}[3]{##3}\VANthebibliography}

\usepackage{graphicx}	
\usepackage{amsmath}
\usepackage{tabularx}
\usepackage{setspace}
\usepackage{natbib}

\usepackage{bm}
\usepackage{amsfonts}
\usepackage{latexsym}
\usepackage{graphicx}
\usepackage{float}
\usepackage{booktabs}
\usepackage{hyperref}
\usepackage{amsmath}
\usepackage{graphicx}
\usepackage{makecell}
\usepackage{enumitem}
\usepackage{setspace}
\usepackage{color}
\usepackage{graphicx}
\usepackage[dvipsnames]{xcolor}
\usepackage{widetext}
\usepackage[misc]{ifsym}
\hypersetup{colorlinks=true,citecolor=blue,linkcolor=red,urlcolor=magenta}
\usepackage{xcolor}
\usepackage{orcidlink}
\usepackage[caption=false]{subfig}
\title[Unveiling the Effects of CEPN Gravity on Cosmic Expansion]{Unveiling the Effects of Coupling Extended Proca-Nuevo Gravity on Cosmic Expansion with Recent Observations}

\author[L.Sudharani et al.]{
L. Sudharani\orcidlink{0000-0002-0860-4584},$^{1}$\thanks{E-mail: sudhaak2694@gmail.com}
N. S. Kavya\orcidlink{0000-0001-8561-130X}$^{1}$\thanks{E-mail: kavya.samak.10@gmail.com}
and V. Venkatesha\orcidlink{0000-0002-2799-2535}$^{1}$\thanks{E-mail: vensmath@gmail.com}\\
$^{1}$Department of P.G. Studies and Research in Mathematics, Kuvempu University, Shankaraghatta, Shivamogga 577451, Karnataka, INDIA
}

\date{Accepted XXX. Received YYY; in original form ZZZ}

\pubyear{2023}

\begin{document}
\label{firstpage}
\pagerange{\pageref{firstpage}--\pageref{lastpage}}
\maketitle

\begin{abstract}
We study Coupling Extended Proca-Nuevo gravity, a non-linear theory extending from dRGT massive gravity with a spin-1 field. This theory is shown to yield reliable, ghost-free cosmological solutions, modeling both the Universe's thermal history and late-time acceleration. By analyzing data from  Dark energy spectroscopic instruments (DESI), Cosmic Chronometer (CCh), Gamma Ray Bursts (GRBs), and Type Ia Supernova (SNeIa), we derive parameter constraints with up to 3$\sigma$ confidence, demonstrating good agreement with observations. Our comparison of $BAO$ data from $WiggleZ$ and $DESI$ highlights its constraining power on the Hubble constant. The analysis of the cosmographic parameter, $q$ shows the statistical compatibility with the recent data. Further, this indicates that Universe's current accelerated expansion aligns with quintessential behavior.

\end{abstract}




\section{Initiation} \label{intro}
The most interesting topic in modern cosmology is the Universe's swift expansion, which poses a challenge to our knowledge of fundamental physics. This phenomenon suggests that the Universe is expanding more quickly with time, unlike what is commonly believed. It was first identified in the late $20^{th}$ century by looking at distant supernovae. Although the simplest
explanation would be the consideration of the cosmological constant, the concordance model of cosmology is highly effective in explaining the evolution of the Universe at both the background and perturbation levels. It is based on general relativity (GR) with a cosmological constant, on the particles of the standard model, and cold dark matter. The two key approaches for creating extended scenarios emerged from the associated challenge relating to the quantum-field-theoretical calculation of its value and the potential for a dynamical nature. However, one approach is to keep GR as the fundamental theory of gravity while taking into account novel and exotic matter that make up the idea of dark energy \citep{Cai:2009zp, Copeland:2006wr}. The second involves creating expanded or modified theories of gravity that, despite having GR as a low-energy limit, generally offer the additional degrees of freedom necessary to propel the acceleration of the dynamic Universe \citep{Capozziello:2011et}.

Nevertheless, according to recent observations of various origins, $\Lambda CDM$ predictions seem to be in tension with the data, as for instance $H_0$ tension, and is observed between two measurements, one from the cosmic microwave background (CMB) temperature and polarization data by the Planck Collaboration \citep{Planck:2018vyg}, which reports $H_0 = 67.37 \pm 0.54$ km s$^{-1}$ Mpc$^{-1}$, and another from local measurements by the Hubble Space Telescope \citep{Riess:2019cxk}, yielding $H_0 = 74.03 \pm 1.42$ km s$^{-1}$ Mpc$^{-1}$. Recent analyses combining gravitational lensing and time-delay effects have reported a significant deviation at $5.3\sigma$ \citep{H0LiCOW:2019pvv}. A further possible source of conflict relates to the measurements of the parameter $\sigma_8$, which measures the gravitational clustering of matter at a scale of $8h^ {-1}Mpc$ based on the amplitude of the linearly developed power spectrum \citep{DiValentino:2020vvd}. However, from the theoretical point of view, $\Lambda$CDM encounters the cosmological constant issue, because GR cannot be approached with a quantum description since it is non-renormalizable. As a result, a lot of effort has gone into developing gravitational modifications, or theories that offer both theoretical and phenomenological benefits while maintaining GR as a limit.

A primary subset of modified gravity theories arises by extending the Einstein-Hilbert Lagrangian with additional terms. This approach leads to a variety of formulations, including $f(R)$, $f(P)$, $f(Q)$, $f(T)$, $f(G)$, and Lovelock gravity theories etc (refer \citep{Starobinsky:1980te, Erices:2019mkd, Lovelock:1971yv, Cai:2015emx,Heisenberg:2023lru}). Further, the scientific community is especially engaged in these gravitational theory classes since they all display complex cosmological features \citep{Bamba:2012cp, Skugoreva:2014ena,Kavya:2024ssu, Mishra:2024oln,Sudharani:2023hss,Naik:2023ykt, Vagnozzi:2019ezj,Pan:2020bur,  Ilyas:2020zcb}. Considering the graviton to be enormous gives rise to an exciting subclass of modified gravity.
Drawing from the framework of massive gravity, researchers proposed the generalized Proca action for a vector field that includes derivative self-interactions, which leads to a theory with only three propagating degrees of freedom. This theory, outlined in \citep{Allys:2015sht, DeFelice:2016yws}, provides a consistent, local description of a massive vector field free from ghost-like instabilities. Consequently, both the background and perturbation-level cosmological solutions were examined \citep{DeFelice:2016yws}. The generalized Proca theory has demonstrated interesting cosmological phenomenology \citep{Heisenberg:2016eld, DeFelice:2016uil, Minamitsuji:2016ydr, BeltranJimenez:2016afo,deFelice:2017paw, Geng:2021jso, DeFelice:2020sdq}.
Proca-Nuevo (PN) theory is a recent variant of Proca theory that has been introduced \citep{deRham:2020yet}. By incorporating non-linear variables for a massive spin-1 field, this theory expands on the conventional Proca framework while preserving crucial consistency restrictions. PN theory can be extended to models providing consistent and ghost-free cosmological answers when combined with gravity. These models include, in particular, late-time self-accelerating phase and hot Big Bang situations. Certain variations of this theory proceed at the speed of light and satisfy all stability and subluminality conditions at the perturbative level. The theory's constraint structure has been thoroughly analyzed and more cosmological solutions have been looked at in \citep{deRham:2021efp,deRham:2023brw, ErrastiDiez:2023gme, Anagnostopoulos:2023pvi, Anagnostopoulos:2021ydo, Gupta:2009kk, Saridakis:2021qxb}. Additionally, studies on the quantum stability of PN interactions \citep{deRham:2021yhr} show that the quantum behaviors of PN and generalized Proca theories are comparable, indicating that they might be particular instances of a larger theoretical framework.

In this paper, we use several data sets to investigate covariant Proca-Nuevo theory in cosmological circumstances. A detailed geometric formulation of integrating Extended Proca-Nuevo theory with gravity and its cosmic background is given in the \sectionautorefname~\ref{epnt}. We present the data sets utilized in \sectionautorefname~\ref{data}, and the approach is described in \sectionautorefname~\ref{method}. The \sectionautorefname~\ref{result} discusses the outcomes, and \sectionautorefname~\ref{conclusion} concludes and offers some thoughts for the future.

\section{Geometric formulation of Coupling Extended Proca-Nuevo with Gravity}\label{epnt}
\subsection{Assessment of Proca-Nuevo and Extended Proca-Nuevo theory}
Let us consider a vector field $\mathcal{V}_{\mu}$ on a flat spacetime with the Minkowski metric $\eta_{\mu\nu}$. The helicity decomposition of vast gravity provides the insight used in the building of Proca-Nuevo (PN) theory. 
 We begin with the equation
 
 \begin{equation}
f_{\mu \nu}[\mathcal{V}] = \eta_{\mu \nu} + 2\frac{\partial_{(\mu} \mathcal{V}_{\nu)}}{\Lambda^2} + \frac{\partial_\mu \mathcal{V}^\rho \, \partial_\nu \mathcal{V}_\rho}{\Lambda^4},
\end{equation}
where $\Lambda$ is an energy scale that will determine the strength of the vector self-interactions.
Similar to the St$\Ddot{u}$ckelberg metric of massive gravity, we must keep in mind that, at this moment, we ignore gravity, therefore $f_{\mu \nu}$ is merely the Lorentz tensor.
For later convenience, we present

\begin{equation}
\phi_r = x^r + \frac{1}{\Lambda^2} \mathcal{V}^r.
\end{equation}
Therefore, $f_{\mu \nu}$ can be written as follows
\begin{equation}
f_{\mu \nu} = \partial_\mu \phi^r \partial_\nu \phi^s\eta_{rs}.
\end{equation}

Although it might initially seem that the dependence of $\phi^r$ on the coordinates $x^r$ indicates a possible violation of Poincar\`e invariance, the actual quantity used as a fundamental component in the Lagrangian is $f_{\mu \nu}$.

In addition, we are going to introduce the tensor 
\begin{equation}
\mathcal{K}^{\mu}_{\,\,\,\nu} = \mathcal{X}^{\mu}_{\,\,\,\nu} - \delta^{\mu}_{\,\,\,\nu},
\end{equation}
and, $\mathcal{X}^{\mu}_{\,\,\,\nu}[\mathcal{V}] = \left( \sqrt{\eta^{-1} f[\mathcal{V}]} \right)^{\mu}_ {\,\,\,\nu}$. Considering the four dimensions, the PN theory for the vector field $\mathcal{V_\mu}$ is then expressed as \citep{deRham:2020yet,deRham:2010kj, Ondo:2013wka,deRham:2011qq,deRham:2014zqa}
\begin{equation}
\mathcal{L}_{PN}[\mathcal{V}] = \Lambda^4 \sum_{n=0}^4 \alpha_n(\mathcal{X}) \mathcal{L}_n[\mathcal{K}].\label{lpn}
\end{equation}
Now, the nth-order PN term is defined as
\begin{equation}
\mathcal{L}_n[\mathcal{K}] = -\frac{1}{(4 - n)!} \epsilon^{\mu_1 \cdots \mu_n \mu_{n+1} \cdots \mu_4} \epsilon_{\nu_1 \cdots \nu_n \mu_{n+1} \cdots \mu_n} \mathcal{K}^{\nu_1}_{\mu_1} \cdots \mathcal{K}^{\nu_n}_{\mu_n}.
\end{equation}
In equation \eqref{lpn}, the PN term is multiplied by a set of coefficients 
$\alpha_n(\mathcal{X})$, which is the arbitrary function of the following equation
\begin{equation}
X = -\frac{1}{2 \Lambda^2} \mathcal{V}^\mu \mathcal{V}_\mu.
\end{equation}
In our analysis, $X$, $\alpha_n$, and $\mathcal{L}_n$ are dimensionless quantities. Notice that $\mathcal{L}_{0}$ is simply a constant, which means that the expression $\alpha_0(X) \mathcal{L}_0 \equiv V(\mathcal{V}^\mu \mathcal{V}_\mu)$ represents the usual potential for the vector field. To ensure that the trivial vacuum state $\langle \mathcal{V}^\mu \rangle = 0$  is physically consistent, $\alpha_{0}$ must include a non-zero quadratic term, specifically, $\alpha_0 \supseteq -\frac{1}{2} \left( \frac{r^2}{\Lambda^4} \right) \mathcal{V}^\mu \mathcal{V}_\mu.$
PN and GP are two distinct ghost-free theories of a massive vector field. An interesting question is whether PN can be extended to include both PN and GP. These models implement the Proca constraint differently, as observed in the null eigenvector (NEV) of their Hessian matrices. In GP theory, the NEV is (1,\~0), indicating that the component \( V_0 \) of the vector field is non-dynamical, similar to the linear theory. Conversely, PN theory uses a field-dependent NEV, as demonstrated in \citep{deRham:2020yet}, where the vector field's constraint is managed differently and is equated as below
\begin{equation}
    V^{PN}_a(\Lambda) = \left(\mathcal{X}^{-1}\right)^{0_\mu} \partial_\mu \phi_a = \left(\mathcal{X}^{-1}\right)^{0_\mu} \left[ \eta_{\mu\nu} + \frac{1}{\Lambda^2} \partial_\nu \mathcal{V}_\mu \right].
\end{equation}
The non-perturbative normalized time-like NEV of the PN Lagrangian, denoted by \( V^{PN}_a\), satisfies the following conditions

\begin{enumerate}
    \item \( \mathcal{H}^{ab} V^{PN}_a = 0\), where \( \mathcal{H}^{ab}\) is the Hessian matrix of time derivatives defined by
    \[
    \mathcal{H}^{ab} = \frac{\partial^2 \mathcal{L}_{PN}}{\partial \dot{\mathcal{V}}_a \partial \dot{\mathcal{V}}_b}.\label{hessian}
    \]
    
    \item The normalization condition
    \[
    \eta_{ab} V^{PN}_a V^{PN}_b = -1,
    \]
    where \( \eta_{ab} \) is the metric tensor.
\end{enumerate}
The PN model is known for its connection to massive gravity, but it is also flexible enough to support additional interactions, any operator that maintains the Hessian invariant can be included without affecting the NEV form. In four dimensions, there are exactly five such operators involving the tensor $\partial_\mu \mathcal{V}_\nu$, called $d_n(X) \mathcal{L}_n[-\mathcal{V}]$, as specified by Hessian matrix. These $\mathcal{L}_n[\partial \mathcal{V}]$ operators are trivial in and of themselves, but when added with a field-dependent coefficient, they can introduce significant effects while remaining trivial in terms of the Hessian; these operators are important to the new derivative interactions in the GP theory (except for $\mathcal{L}_4$). Notably, operators not based on elementary symmetric polynomials of $\partial^\mu \mathcal{V}_\nu$ generally affect the Hessian and thus cannot be added as easily.

Here, some redundancies arise from the construction described
\begin{itemize}
    \item $\mathcal{L}_0[\partial \mathcal{V}]$ is constant and so its coefficient can be absorbed into $\alpha_0$, affecting the non-derivative potential.
    \item Only three of the four remaining terms are linearly independent from the PN operators.
    \item$ f(X)\mathcal{L}_4[\partial \mathcal{V}]$ is always a total derivative for any function $f$, making this term redundant.
\end{itemize}
The above properties hold only in flat spacetime and do not apply to curved backgrounds. Thus, when developing a covariant theory, all four GP terms ($\mathcal{L}_1[\partial \mathcal{V}]$ through $\mathcal{L}_4[\partial \mathcal{V}]$) must be considered.

Having these factors, we present the following Lagrangian

\begin{equation}
\mathcal{L}_{\text{EPN}} = \tilde{\Lambda}^4 \sum_{n=0}^4 \alpha_n(\tilde{\mathcal{X}}) \mathcal{L}_n[\tilde{\mathcal{K}}[\mathcal{V}]] + \Lambda^4 \sum_{n=1}^4 d_n(X)\frac{\mathcal{L}_n[\partial \mathcal{V}]}{\Lambda^{2n}}, 
\end{equation}

which is termed as the “Extended Proca-Nuevo” (EPN) theory. This Lagrangian tribute four additional arbitrary functions, \(d_n(X)\), beyond the original functions \(\alpha_n\) in four dimensions. It is important to note that we have permitted the two sets of operators to appear at different scales, \(\Lambda\) and \(\tilde{\Lambda}\), and that \(\tilde{\mathcal{K}}\) and \(\tilde{\mathcal{X}}\) are the quantities defined earlier, but scaled by \(\tilde{\Lambda}\).

\subsection{Cosmological background to the Covariant Extended Proca-Nuevo theory with Gravity}

The search for the most general theory of a self-interacting massive spin-1 and interacting effective field theories involving fields of different spins are fascinating subjects that have seen significant advancements in the last ten years. The integration of these efficient field theories into a gravitational framework, especially for astrophysical and cosmological applications, is an interesting task as it advances the continuous endeavor to categorize practical extensions of GR. 

This section presents the Covariant Extended Proca-Nuevo (CEPN) theory coupled with gravity \citep{deRham:2021efp}. The action for the CEPN theory is given by

\begin{equation}\label{action}
     \mathcal{S}=\int d^4 x\sqrt{-g}\left(\frac{M^2_{Pl}}{2}R+\mathcal{L}_{EPN}+\mathcal{L}_M\right).
 \end{equation}

R and $\mathcal{L}_M$ represent Ricci scalar and standard matter Lagrangian. Now we can define massive spin-1 Lagrangian as given below

\begin{equation}\label{spin}
  \mathcal{L}_{EPN}=-\frac{1}{4}\mathcal{F}^{\mu\nu}\mathcal{F}_{\mu\nu}+\Lambda^4\left(\mathcal{L}_0+\mathcal{L}_1+\mathcal{L}_2+\mathcal{L}_3\right),  
\end{equation}
where,

\begin{gather}
\mathcal{L}_0 = \alpha_0(X), \\
\mathcal{L}_1 = \alpha_1(X)\mathcal{L}_1[\mathcal{K}] + d_1(X)\mathcal{L}_1\frac{[\nabla \mathcal{V}]}{\Lambda^2}, \\
\mathcal{L}_2 = [\alpha_2(X) + d_2(X)] \frac{R}{ \Lambda^2} + \alpha_{2,X}(X)\mathcal{L}_2[\mathcal{K}] + d_{2,X}(X) \mathcal{L}_2\frac{[\nabla \mathcal{V}]}{\Lambda^4}, \\
\begin{split}
\mathcal{L}_{3} = \left[\alpha_3(X)K^{\mu\nu} + d_3(X)\frac{\nabla_{\mu}\mathcal{V}_{\nu}}{\Lambda^2}\right] \frac{G^{\mu\nu}}{\Lambda^2}- \frac{1}{6}\alpha_{3,X}(X)\mathcal{L}_3[\mathcal{K}] \\- \frac{1}{6}d_{3,X}(X)\frac{\mathcal{L}_3[\nabla \mathcal{V}]}{\Lambda^6}.     
\end{split}
\end{gather}

Non-minimal coupling terms, proportional to $R$ in $\mathcal{L}_2$ and the Einstein tensor $G_{\mu\nu}$ in $\mathcal{L}_3$, are included in the Lagrangian. Along with the non-minimal couplings question, this Lagrangian does exclude the $\mathcal{L}_4$ term that existed in flat spacetime.

Now, the FLRW metric is our main emphasis
\begin{equation}
ds^2 = -N^2(t)dt^2 + \mathcal{A}^2(t) \delta_{ij} {dx^i}{dx^j},
\end{equation}
here, $A(t)$ is the Universe scale factor and we may set $N=1$. Further, the vector field profile is defined below
\begin{equation}
    \mathcal{V}_\mu dx^{\mu}=-\phi(t) dt.
    \end{equation}
Now we yield a modified Friedmann equation by varying the action concerning the lapse \eqref{action}
\begin{gather}
H^2 = \frac{1}{3M_{Pl}^2}(\rho_m +\tilde{ \rho}_{EPN}),\label{fdmn1} \,\,\text{and}\\
\dot{H} + H^2 = -\frac{1}{6M_{Pl}^2}(\rho_m + \tilde{\rho}_{EPN} + 3p_m + 3\tilde{p}_{EPN}).\label{fdmn2}
\end{gather}
In the above equation, $\rho_m$ and $p_m$ respectively represent the energy density and the pressure of the matter fluid.
However, variation with respect to $\phi(t)$ gives the following

\begin{equation}
    \alpha_{0,X} + 3(\alpha_{1,X} + d_{1,X})\frac{H\phi}{\Lambda^2} = 0.\label{scalar }
\end{equation}
An effective dark energy sector with energy density and pressure are given by

\begin{gather}
\tilde{\rho}_{EPN} \equiv \Lambda^4 \left( -\alpha_0 +\alpha_{0,X} \frac{\phi^2}{\Lambda^2} + 3(\alpha_{1,X} + d_{1,X})\frac{H \phi^3}{\Lambda^4 }  \right),\,\,\,\, \text{and}\\
\tilde{p}_{EPN} \equiv \Lambda^4 \left( \alpha_0 - (\alpha_{1,X} + d_{1,X}) \frac{\phi^2 \dot{\phi}}{\Lambda^4} \right).
\end{gather}

The scalar field equation \eqref{scalar } is non-dynamical, it is just a constraint that imposes an algebraic relationship between $H$ and $\phi$. This is due to action integral \eqref{action}, therefore, the resulting Friedmann equations depend only on the Hubble function and not on the vector field. Therefore, the effective dark energy density and pressure can be shown as

\begin{gather}
\tilde{\rho}_{EPN} = \Lambda^4  \frac{c_m y^{2/3}}{2}  \left( \frac{\Lambda^4}{M_{Pl}^2 H^2} \right)^{1/3}, \,\,\,\, \text{and}\label{repn} \\
\tilde{p}_{EPN} = 3M_{Pl}^2 H^2 \left( -1 - \frac{2\dot{H}}{3H^2} \right).
\end{gather}
Here, $c_m\equiv\frac{m^2M^2_{pl}}{\Lambda^4}\sim 1$ and $y=4\sqrt{\frac{6}{c_m}}$ (refer \citep{deRham:2021efp}). Moreover, we consider the density parameters $\Omega_M\equiv \rho_M/3M^2_{Pl}H^2$ and $\Omega_{EPN}\equiv \tilde{\rho}_{EPN}/3M^2_{Pl}H^2$. Now using the first Friedmann equation \eqref{fdmn1} at $z=0$, then equation \eqref{repn} become
\begin{equation}
\frac{\Lambda^{16/3}c_m y^{2/3}}{6M^{8/3}_{Pl}H^{8/3}_0}-(1-\Omega_{m_0})=0.\label{simplified}
\end{equation}
Upon substituting \eqref{simplified} into \eqref{repn} and eliminating $c_m y^{2/3}$ and $\Lambda$, we obtain
\begin{equation}
    \tilde{\rho}_{EPN} = 3M^2_{Pl}(1-\Omega_{m_0})H^{8/3}_0 H^{-2/3}.
\end{equation}
Utilizing all these equations in \eqref{fdmn1}, we obtain an algebraic equation defined as follows
\begin{equation}\label{eq:H}
    H^2-H^2_0\Omega_{m_0}(1+z)^3-(1-\Omega_{m_0})H^{8/3}_{0}H^{-2/3}=0,
\end{equation}
where $H$ is the Hubble function. As presented in \citep{Anagnostopoulos:2023pvi}, the deviation from $\Lambda$CDM is evident from the presence of $(H_0 H)^{-2/3}$ in the in the third term of LHS of equation \eqref{eq:H}. The powered-Hubble parameter $H^{-2/3}$ becomes less prominent for infinitely large redshifts. When considering the present case, \eqref{eq:H} mimics $\Lambda$CDM scenario. Hence $H^{-2/3}$ becomes significant in analyzing the dynamics of the CEPN Universe at intermediating stages, where one can witness a shift from the standard cosmological model. To this end, one cannot ignore its presence in the equation \eqref{eq:H}. However, when this term is considered, it is challenging to find the exact solution of the Hubble parameter. Thus, in our analysis, we numerically account for this to provide a more detailed assessment. 
\begin{figure*}
    \centering
    {\includegraphics[width=0.4\linewidth]{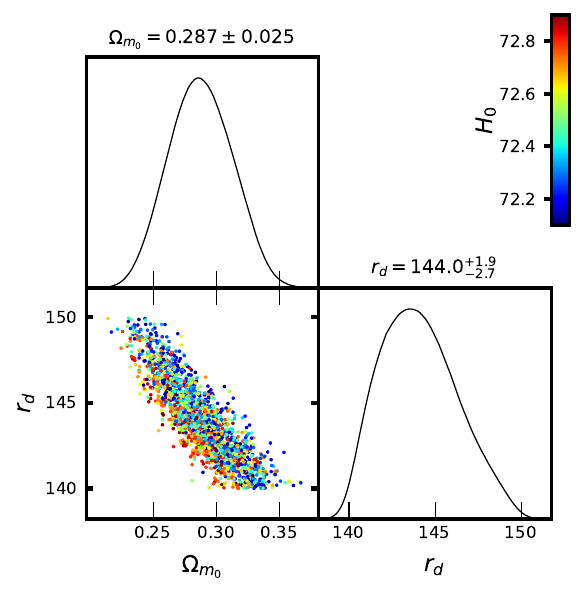}}
    {\includegraphics[width=0.4\linewidth]{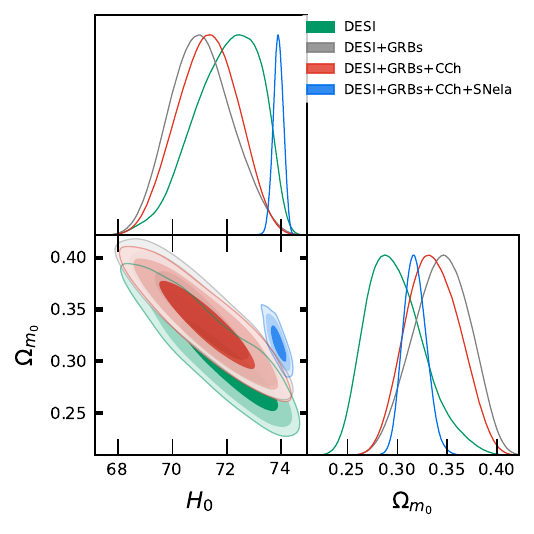}}
    \caption{ Left: The contour plot for DESI data illustrating the constraints on the Hubble parameter $H_0$, the sound horizon $r_d$, and $\Omega_{m_0}$. Right: A contour plot showing the model parameters $H_0$ and $\Omega_{m_0}$ obtained through $\chi^2$ analysis for the current model. This plot illustrates the results of a combined analysis of various datasets, with confidence levels up to $3\sigma$. For DESI data, $r_d$ is fixed which is the mean value obtained in the Left panel.  }
 \label{fig:combined}
\end{figure*}

    \begin{figure*}
\centering
    \includegraphics[width=0.33\linewidth]{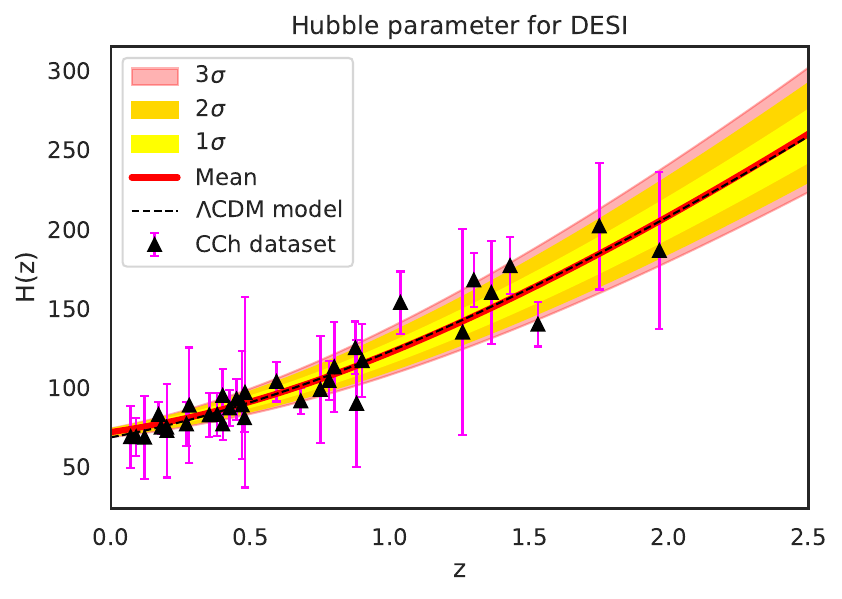}
    \includegraphics[width=0.32\linewidth]{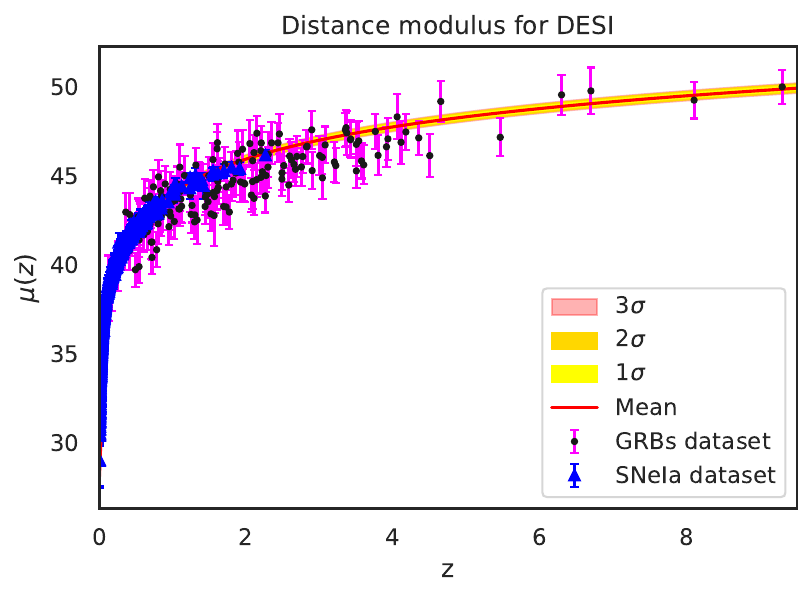}
    \includegraphics[width=0.33\linewidth]{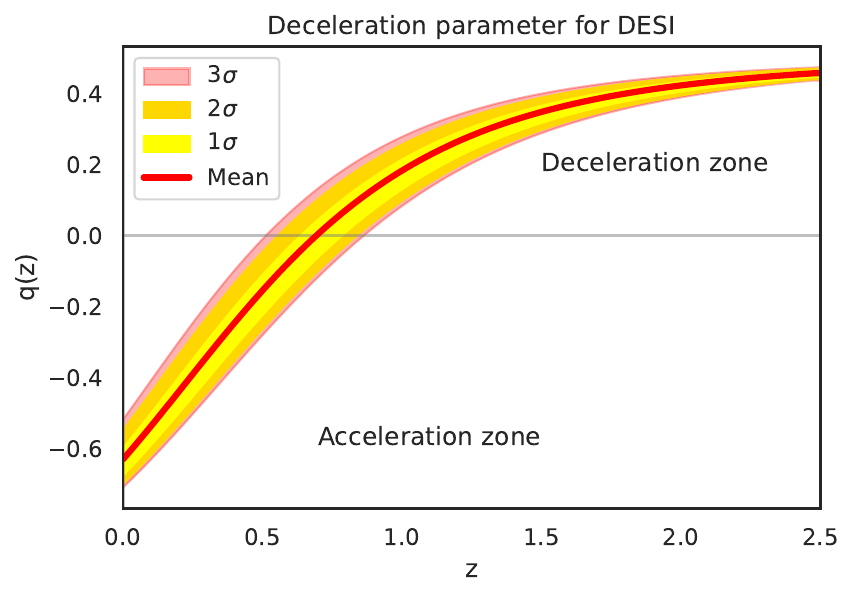}\\
    \includegraphics[width=0.33\linewidth]{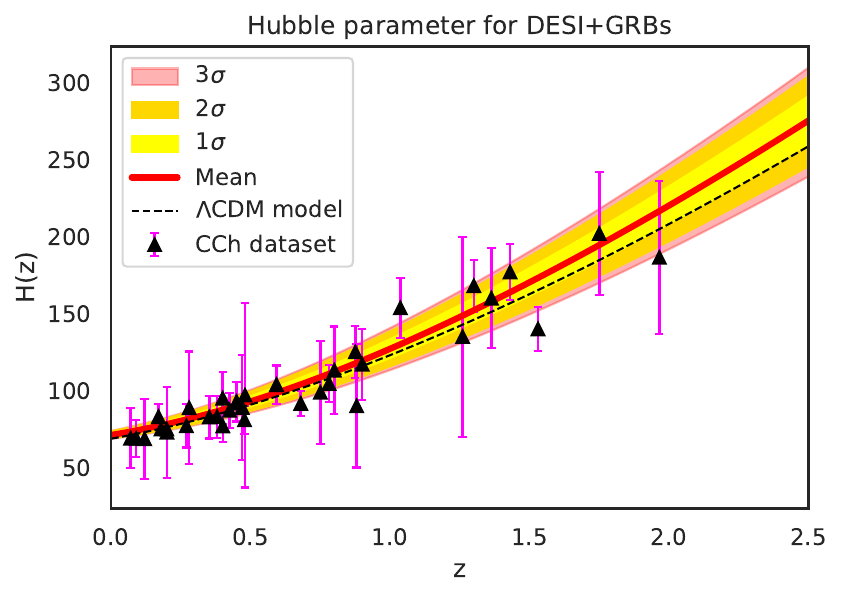}
    \includegraphics[width=0.32\linewidth]{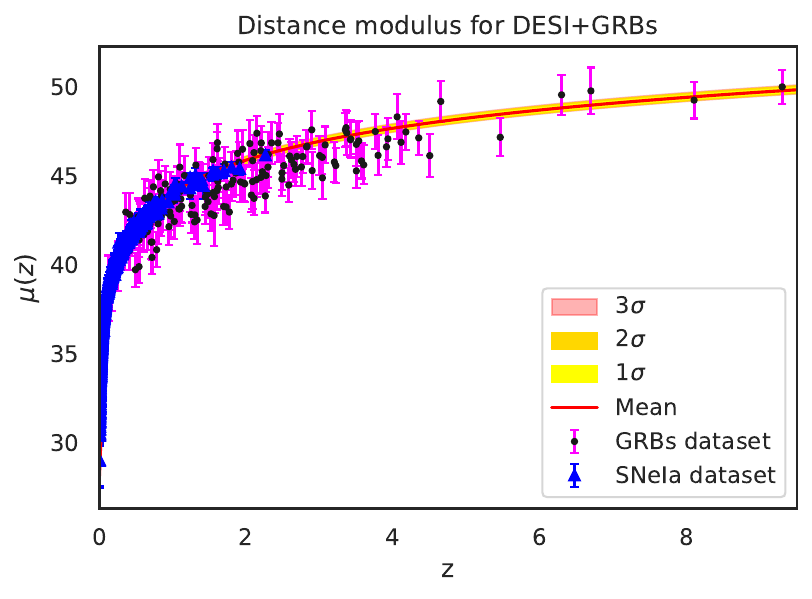}
    \includegraphics[width=0.33\linewidth]{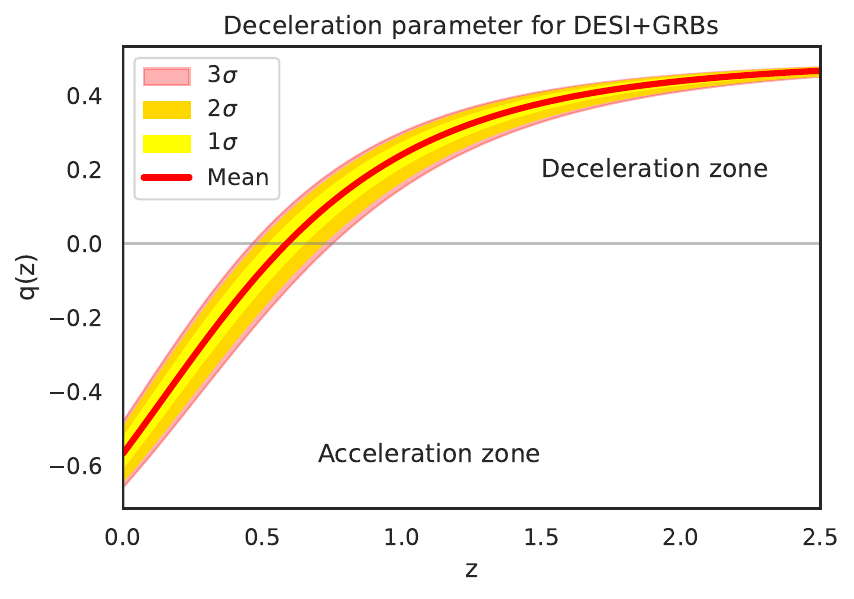}\\
    \includegraphics[width=0.33\linewidth]{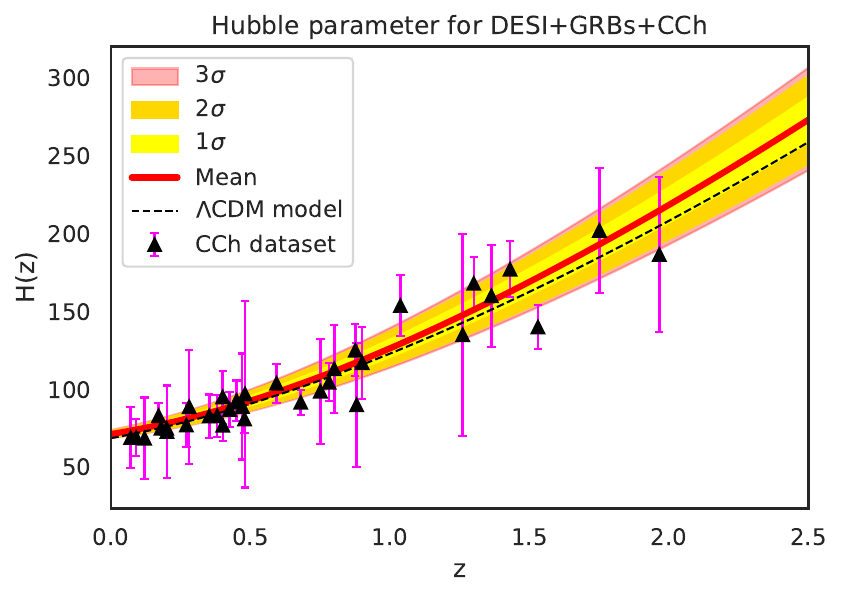}
    \includegraphics[width=0.32\linewidth]{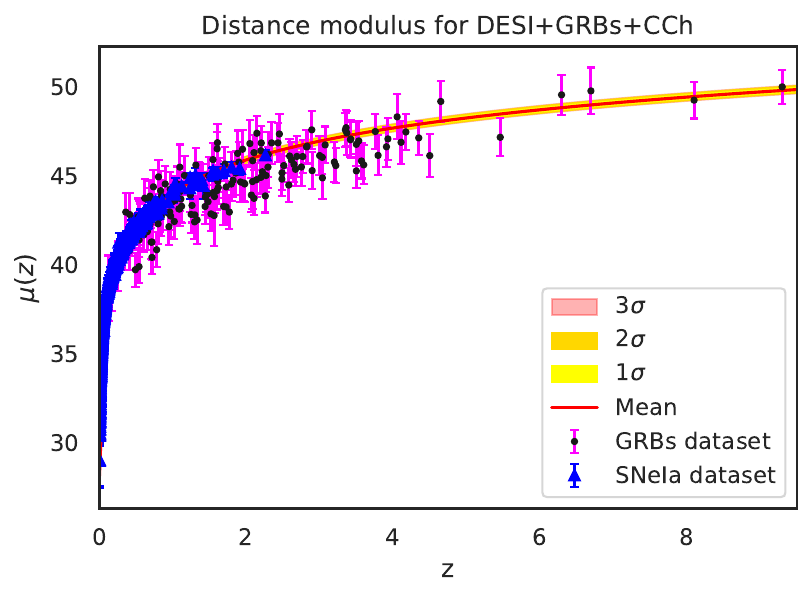}
    \includegraphics[width=0.33\linewidth]{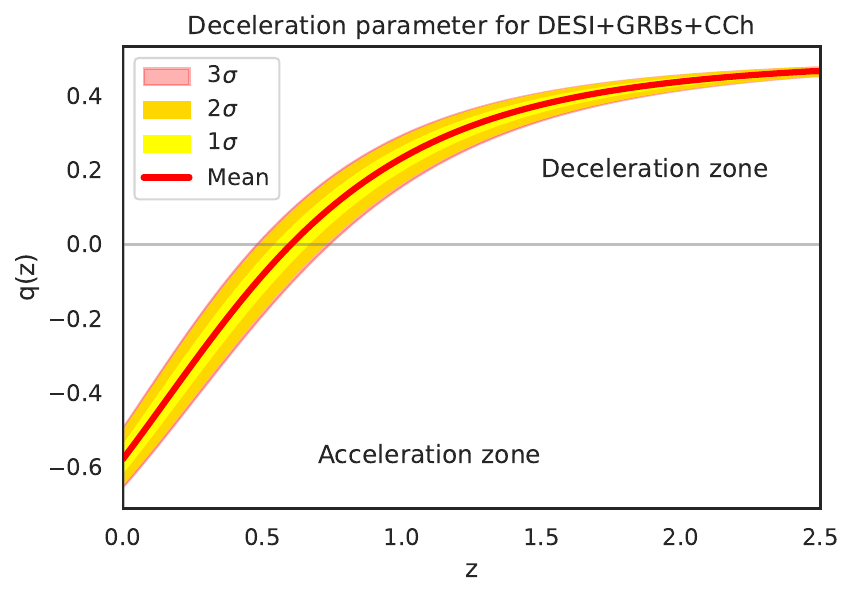}\\
    \includegraphics[width=0.33\linewidth]{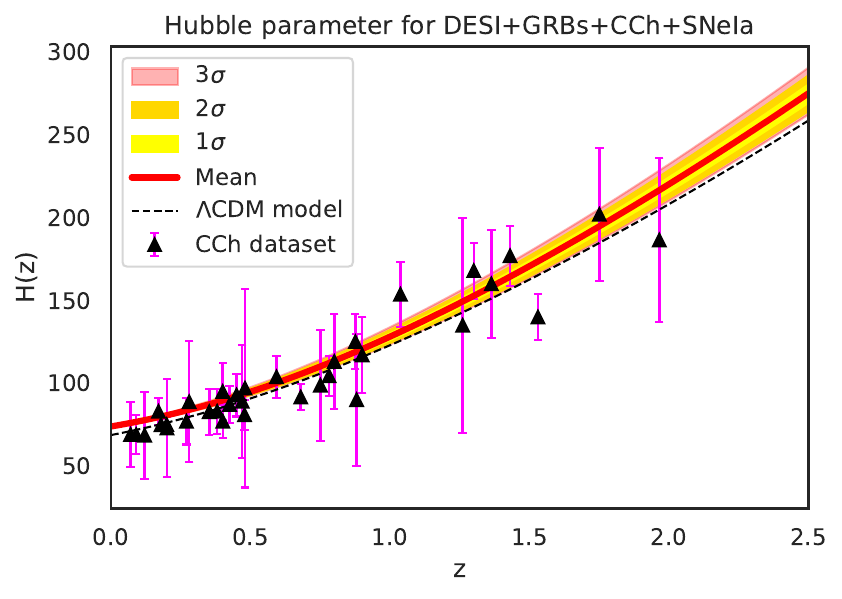}
    \includegraphics[width=0.32\linewidth]{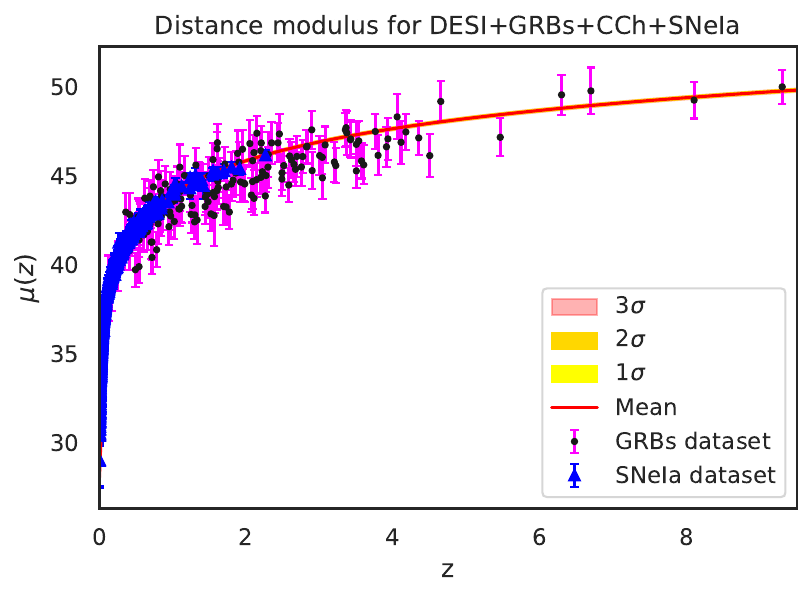}
    \includegraphics[width=0.33\linewidth]{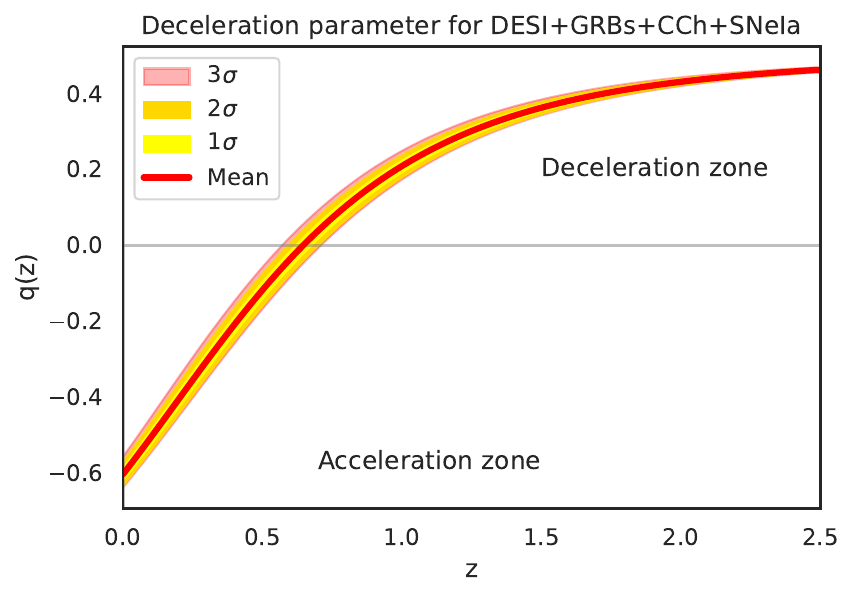}
    \caption{Cosmographic parameter analysis: The first column displays $H(z)$ data with theoretical predictions (red line) and shaded confidence regions. The second column, with high-confidence error bars, shows the distance modulus $\mu(z)$ for the 1701 SNeIa and 162 GRBs data points. The third column illustrates the deceleration parameter $q$.}
    \label{fig:cosmographic_parameters1}
    \end{figure*}

     \begin{figure*}
\centering
    \includegraphics[width=0.33\linewidth]{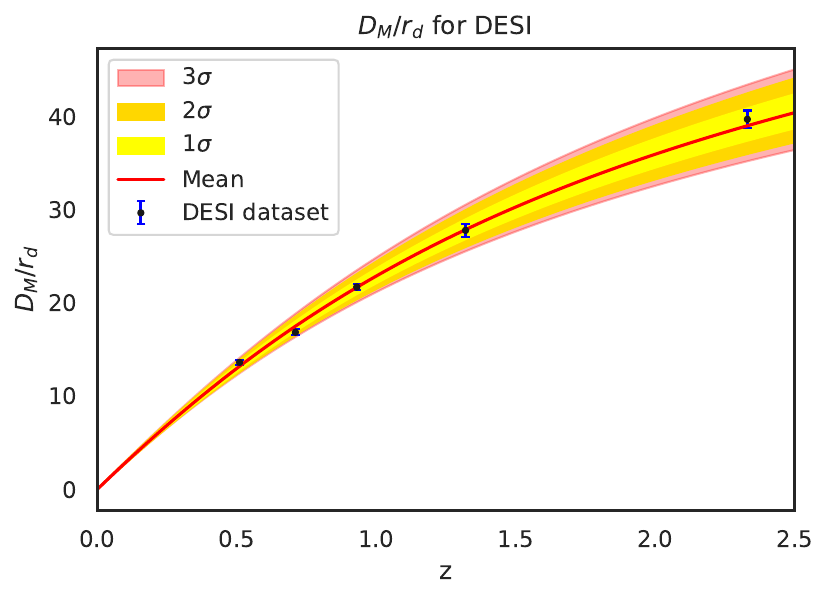}
    \includegraphics[width=0.33\linewidth]{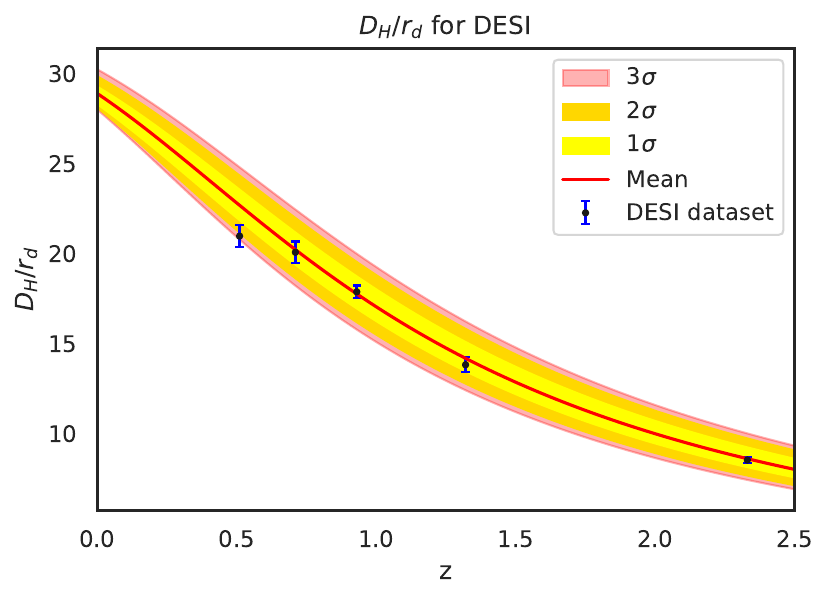}
    \includegraphics[width=0.33\linewidth]{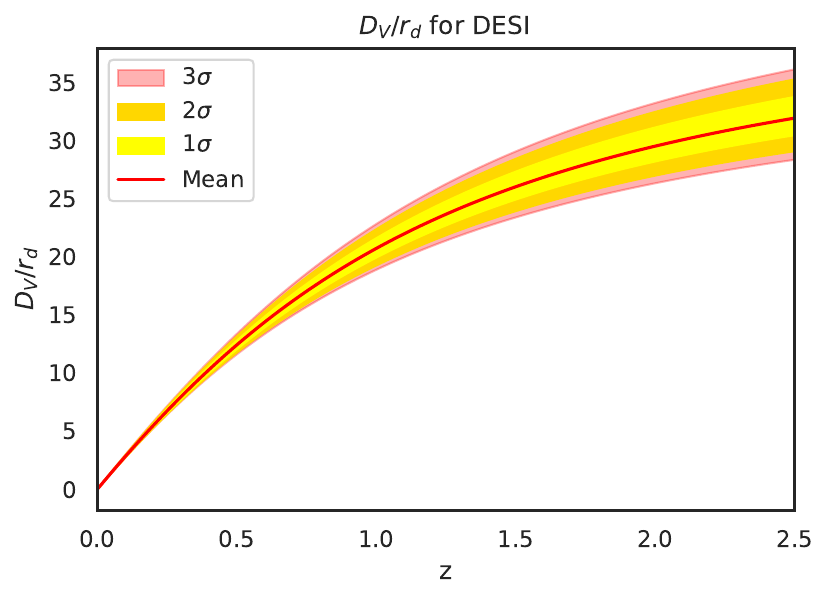}\\
    \includegraphics[width=0.33\linewidth]{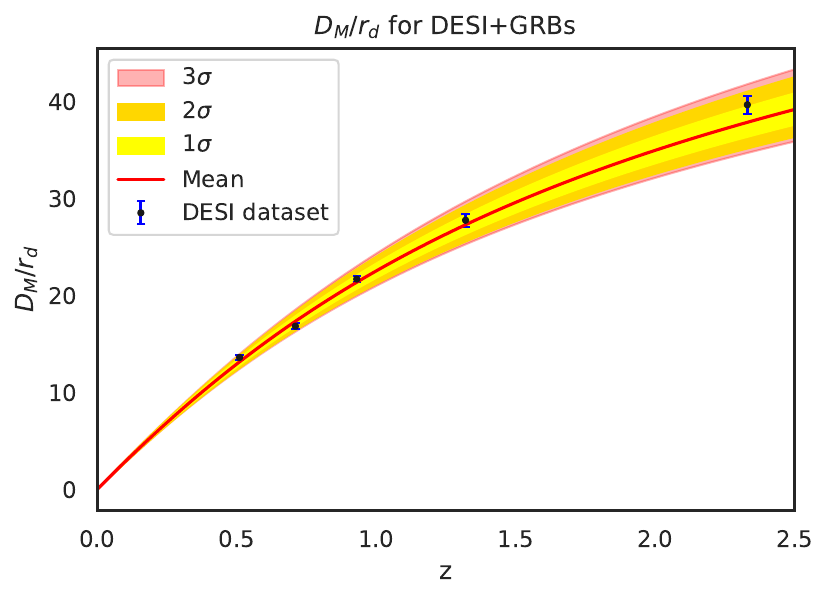}
    \includegraphics[width=0.33\linewidth]{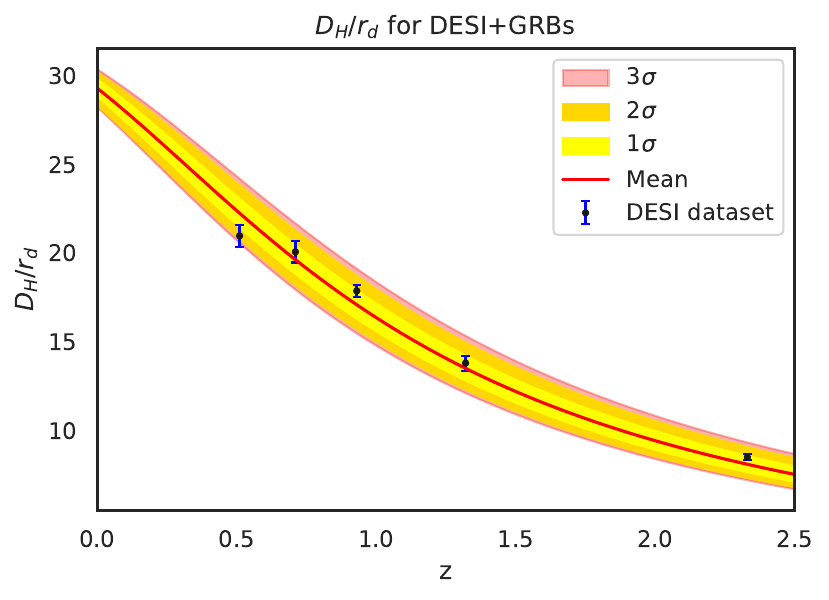}
    \includegraphics[width=0.33\linewidth]{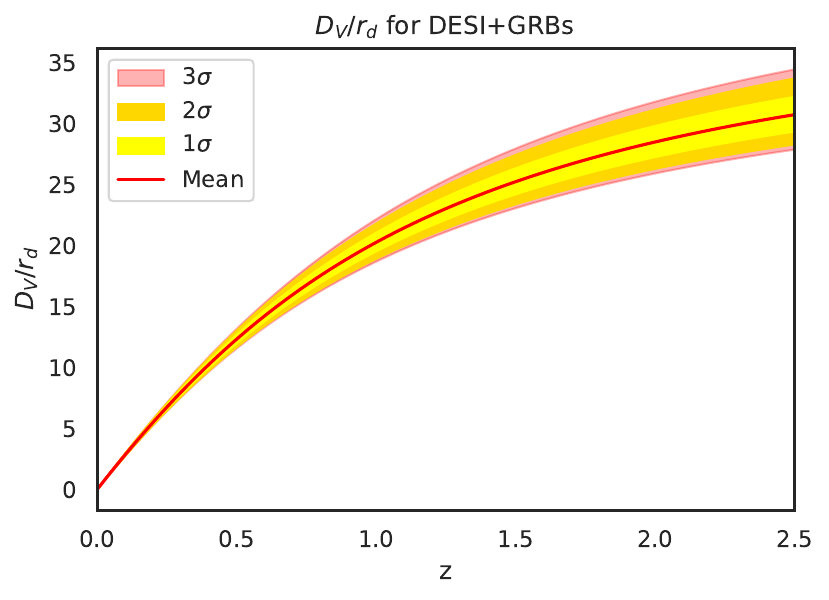}\\
    \includegraphics[width=0.33\linewidth]{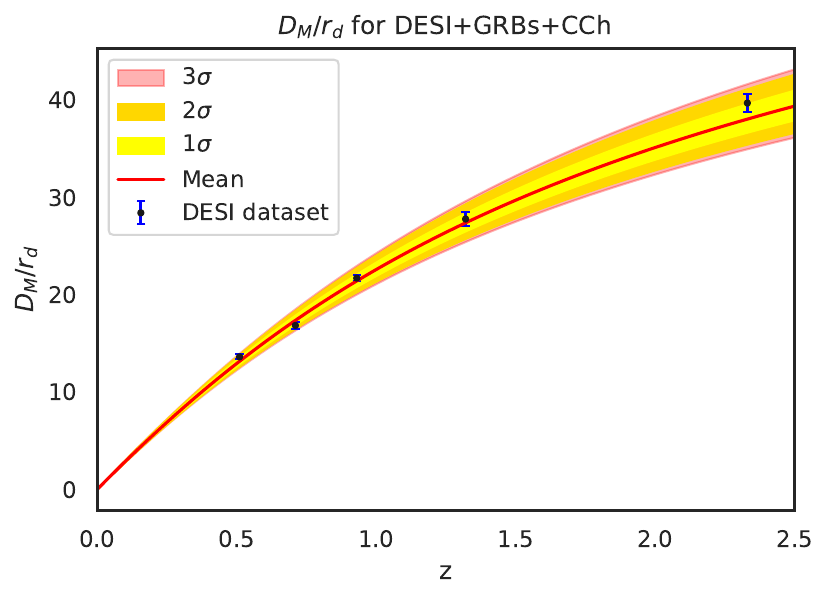}
    \includegraphics[width=0.33\linewidth]{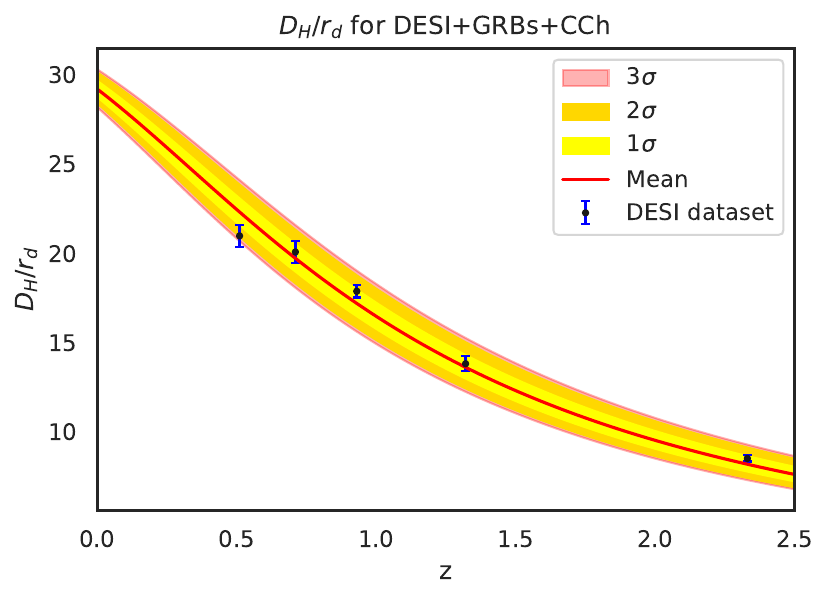}
    \includegraphics[width=0.33\linewidth]{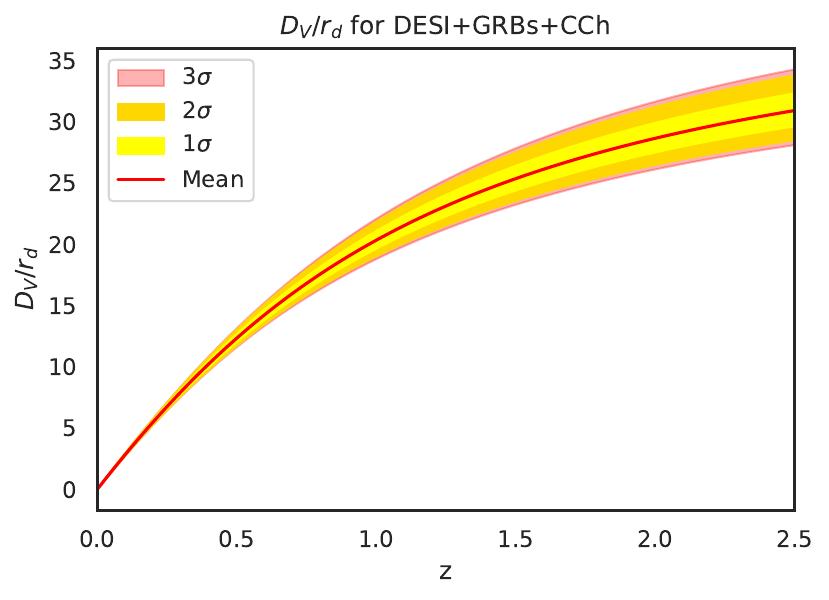}\\
    \includegraphics[width=0.33\linewidth]{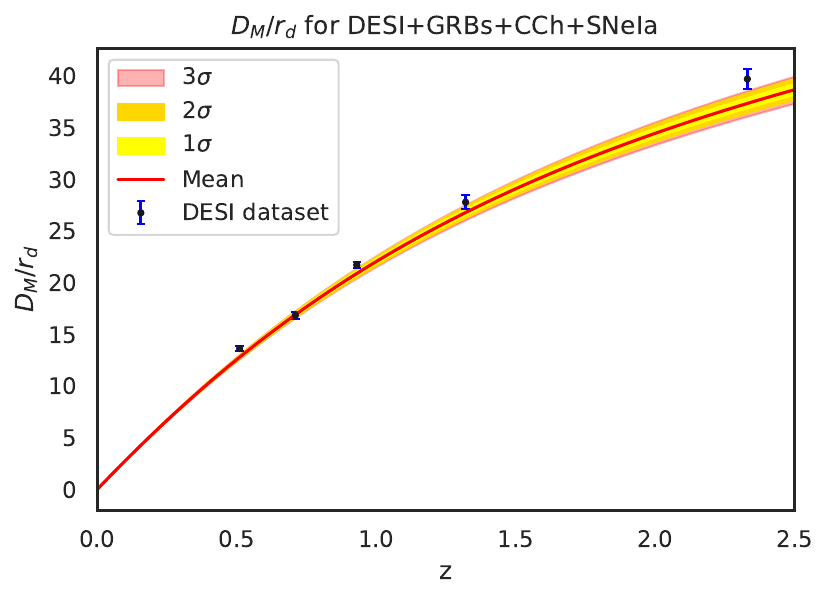}
    \includegraphics[width=0.33\linewidth]{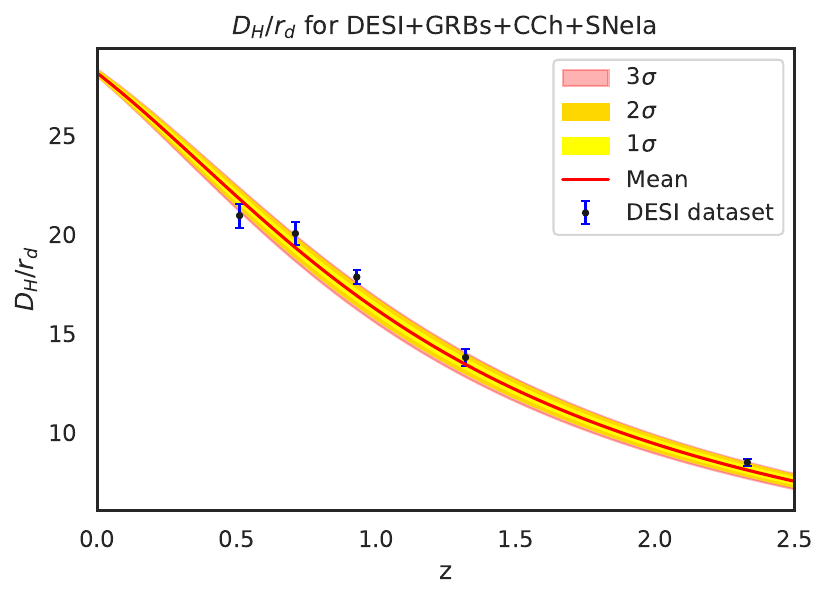}
    \includegraphics[width=0.33\linewidth]{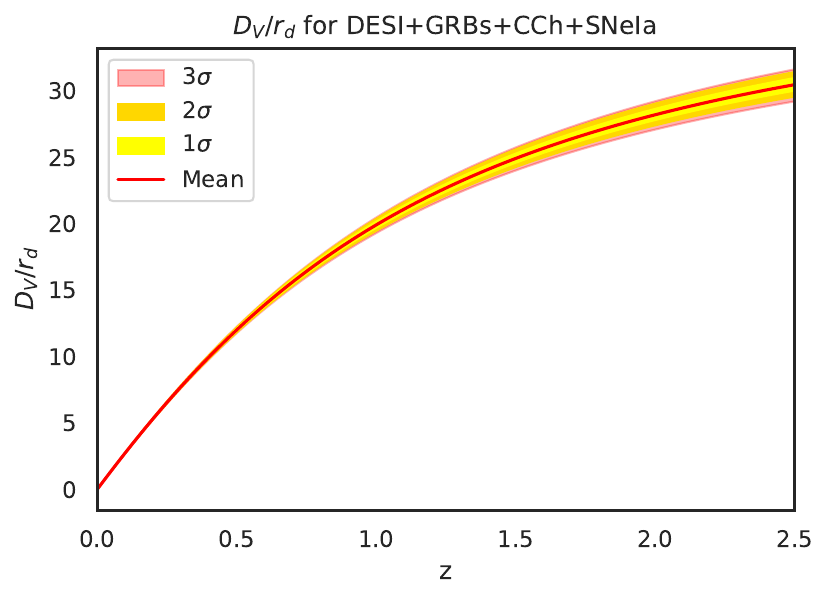}
    \caption{Cosmographic parameter analysis: The plot features mean curves for the comoving angular diameter distance $D_M$, Hubble distance $D_H$, and volume distance $D_V$, normalized by the sound horizon parameter $r_d$ and plotted against redshift. The red lines representing theoretical predictions from the CEPN model align closely with the $DESI$ data error bars, highlighting the model's accuracy and consistency with observational data.}
    \label{fig:cosmographic_parameters2}
    \end{figure*}

\begin{figure*}
    \centering
    {\includegraphics[width=0.4\linewidth]{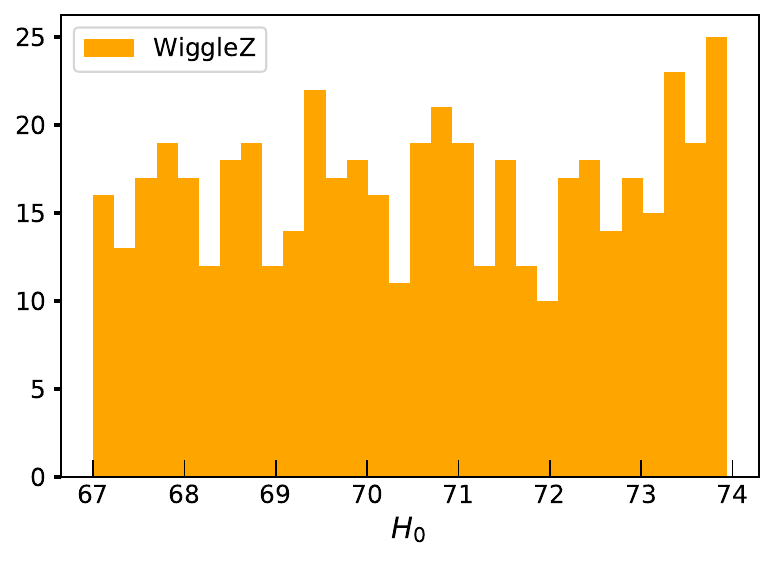}}
    {\includegraphics[width=0.4\linewidth]{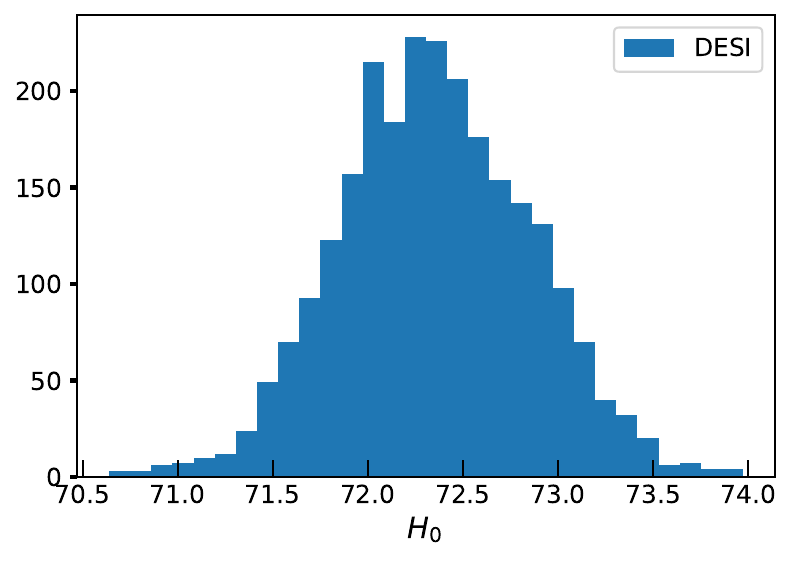}}
    \caption{Comparison of the constraining power of $BAO$ data on the model parameter $H_0$ from the $WiggleZ$ and $DESI$ surveys, based on the distribution of the curves. }
    \label{fig:histogram}
\end{figure*} 

\section{Data}\label{data}
This section presents the datasets used to investigate the dark energy model obtained in the previous section.

\subsection{DESI BAO}\label{desi}
As a consequence of acoustic density waves in the primordial plasma of the early cosmos, changes in the density of baryonic matter occurs and is known as baryonic acoustic oscillations (BAO). This source offers information and correlations for both the distance variable $(D_H/r_d)$ and the comoving distance $(D_M/r_d)$ throughout the drag period. The drag epoch, denoted by the symbol $z_d$, and the radius of the sound horizon $r_d$ is the farthest distance sound waves may have traveled between the Big Bang and the time at which baryons decoupled.
In cases where there is a poor signal-to-noise ratio, the averaged quantity $ D_V/r_d$ is employed. It is noteworthy that the BAO signal from galaxy clustering has been detected by the ``Dark energy spectroscopic instrument (DESI)" employing several tracers of matter, including galaxies, quasars, and Laman-$\alpha$ forests \citep{DESI:2024uvr, DESI:2024lzq}. It is only possible to ascertain the whole value of $r_dH_0$ when utilizing solely DESI BAO data. However, we can differentiate between $r_d$ and $H_0$ independently when we combine DESI BAO data with other observational datasets.

 To do this, we used observations from DESI BAO Data Release 1 in the redshift range of $0.3$ to $2.33$ (see Tab. I of \citep{Pourojaghi:2024tmw}).
There are two isotropic BAO datasets, the Quasar (QSO) at an effective redshift of $z_{eff} = 1.49$ and Bright Galaxy Survey (BGS) at $z_{eff} = 0.30$.
Five data points are also included in the anisotropic BAO datasets: Luminous Red Galaxy (LRG) at $z_{eff}$ of $0.51$ and $0.71$, LRG+ELG at $z_{eff} = 0.93$, Emission Line Galaxies (ELG) at $z_{eff} = 1.32$, and Lyman-$\alpha$ quasars (Lya QSOs) at $z_{eff} = 2.33$. 

At the baryon drag epoch, $r_d$, the sound horizon can be identified as
\begin{equation}
    r_d=\int^{\infty}_{z_d}\frac{c_s(z)}{H(z)}dz,
\end{equation}
where $c_s(z)$ is the fluid sound speed of baryons-photons. The transverse comoving distance to the tracers at each redshift bin can be calculated using this physical scale as a ruler. In a flat Universe, this distance is given by
\begin{equation}
    D_M(z)=c\int^z_0 \frac{d\xi}{H(\xi)},
\end{equation}
as well as the Hubble factor at the tracer redshift, which can be used to calculate a distance
\begin{equation}
    D_H(z)=\frac{c}{H(z)}.
\end{equation}
Here $c$ is the speed of light, and $H(z)$ is the Hubble parameter. Additionally, for the rest of this work, we simply refer to this dataset as DESI.

\subsection{CCh}\label{cc}
The cosmic chronometers (CCh) method is a conceptually simple technique to measure the Hubble parameter as a function of redshift, $H(z)$, independent of the cosmological model adopted \citep{Jimenez:2001gg}. While redshift measurements can achieve high precision ($\delta z/z\leq0.001$) through spectroscopy of extragalactic objects, the main challenge lies in accurately estimating the differential age evolution $dt$. This challenge necessitates the use of a ``chronometer". The primary advantage of the CCh approach is its ability to directly
estimate the expansion history of the Universe without the need for any prior cosmological assumptions. We applied methods from \citep{Moresco:2020fbm} as well as research that takes into account statistical and systematic elements in our investigations. Nonetheless, we employed data retrieved from CCh at redshifts between $0.07$ and $1.26$ (see Table 1. in \cite{Sudharani:2023vhv}).

\subsection{GRBs}\label{grb}
The extraordinarily powerful and luminous events known as Gamma-Ray Bursts (GRBs) were discovered by the Vela satellites more than fifty years ago \citep{Klebesadel:1973iq}. Up to very high redshifts, these sources have been observed further; the highest redshifts are z = 8.2 \citep{Tanvir:2009ex} and z = 9.4 \citep{Cucchiara:2011pj}. They are an important tool to shed new light on the significant cosmological tensions that exist today and to overcome the knowledge gap on the evolution of the Universe between the farthest type Ia supernovae and the Cosmic Microwave Background Radiation. To distinguish between many potential origins, a classification of GRBs according to their measured light curves is essential. Historically, GRBs have been classified into two primary classes: Short GRBs (SGRBs) and Long GRBs (LGRBs), depending on how long the prompt emission lasted. For an extensive review of the GRBs prompt correlations, we refer \cite{Dainotti:2017fhk, Dainotti:2023bwq, Dainotti:2022rfz, Dainotti:2017fem}. Referring to Table 5 of the publication \citep{Demianski:2016zxi}, we analyzed a sample of 162 LGRBs, whose redshift distribution ranges widely, with values between $0.03 \leq z \leq 9.3$.

\subsection{Type Ia Supernova}\label{sneia}
Supernovae of Type Ia (SNeIa) are `standard candles' because of their intrinsic luminosity. These occur when a star is nearing the end of its life and produces a tremendous explosion that disperses stellar material across space. Researchers can determine their precise distances and provide a different method of measuring the expansion history of the Universe by analyzing their apparent brightness and redshift.

In this work, we analyze one of the distinct samples of SNeIa, which includes the Pantheon+ sample \citep{Scolnic:2021amr}. This sample represents the most recent collection of spectroscopically confirmed SNeIa, examined across the entire redshift range between the value of $0$ and $ 2.3$. Moreover, the Pantheon+ compilation contains $1701$ SN light curves, $77$ of which originate from galaxies that host Cepheids in the low redshift interval $0.001\leq z \leq 2.2613.$ The compilation includes $18$ distinct investigations \citep{Scolnic:2021amr, Pan-STARRS1:2017jku}.
We introduce the following vector
\[
Q_i =
\begin{cases} 
m_i - M - \mu_i & \text{if } i \in \text{Cepheid hosts} \\
m_i - M - \mu_{\text{model}}(z_i) & \text{otherwise}
\end{cases}
\]
to reconcile the degeneracy between the Hubble constant $H_0$ and the absolute magnitude $M$ of the SNeIa.
where the apparent magnitude and distance modulus of the $i^{th}$ SNeIa are denoted by $m_i$ and $m_{i}-M$, respectively.  

\section{Methodology}\label{method}
This section presents the methodology we employed in this work to do inference of cosmological parameters utilizing several data sets. The Markov Chain Monte Carlo (MCMC) technique is the methodology used in our work. MCMC is a probabilistic technique that generates a series of samples, each of which is a set of parameter values, in order to explore the parameter space.
The method's central tenet is the Markov property, which states that the sequence's subsequent state depends only on the one before it or a current state. When MCMC is applied to cosmological parameter inference, it is frequently used to sample the posterior distribution of parameters given observational data. Essentially, it explores the parameter space by generating a chain of samples, the density of which reflects the posterior distribution, and then analyzing the chain to determine the most probable values and uncertainties for cosmological parameters. From a Bayesian perspective, parameter estimation treats probability as a measure of belief. Parameters are seen as random variables that are updated based on data and prior beliefs. Initially, a prior distribution is set, and the data refines this through an iterative process to produce a posterior distribution. This posterior reflects all the knowledge gained, integrating prior beliefs with new evidence from the data.

From the likelihood and the prior, we can derive the posterior distribution using the Bayes theorem
\begin{equation}
    P(\Theta \mid D) = \frac{P(D \mid \Theta) \cdot \pi(\Theta)}{P(D)} = \frac{P(D \mid \Theta) \cdot \pi(\Theta)}{\int P(D \mid \Theta) \cdot \pi(\Theta) \, d\Theta}.
\end{equation}
However, Bayesian statistics does inference using the rule of probability directly and is based on a single tool, Bayes' theorem, which finds the posterior density of the parameters provided by the data. It combines both the prior information we have in the prior $g(\theta_1,...,\theta_p)$ and the information about the parameters contained in the observed data given likelihood $f(y_1,...,y_,|\theta_1,...,\theta_p)$. 

The primary problem with estimating Bayesian parameters is that it is generally impossible to obtain the posterior distribution analytically, and even numerical analysis is frequently not feasible. Through the clever utilization of MCMC sampling, this enduring problem was solved, marking a major advancement in statistics throughout the 20th century. When applied to prolonged sequences, MCMC yields a sequence of parameter sets (Markov chain) whose empirical distribution approaches or converges to the posterior distribution.

As the sequence lengthens, the empirical distribution of the parameter sample generated by MCMC sampling converges to the true posterior distribution, providing a sophisticated solution for evaluating model parameters, especially when the corresponding posterior distribution cannot be accessed analytically. As a result, any question concerning the posterior parameter distribution can theoretically be answered simply by looking at the corresponding Markov chain.

Monte Carlo methods are designed to approximate a target density \( p(x) \) where \( x \in X \) and \( X \) is a high-dimensional space. This is achieved by generating a set of independent and identically distributed samples \( \{ x^{(i)} \}_{i=1}^{N} \). These samples are then used to estimate integrals or maxima of the target function. While basic sampling techniques can handle simple forms of \( p(x) \), more advanced methods, such as MCMC, are required for complex real-world problems.

One state in a Markov chain depends only on the one that came before it, a sequence of states produced by a stochastic process known as the ``Markov property",
\begin{equation}
    P\left(x^{(i)} \mid x^{(i-1)}, \ldots, x^{(1)}\right) = P\left(x^{(i)} \mid x^{(i-1)}\right)\,\, \text{and}
\end{equation}

\begin{equation}
    T\left(x^{(i)} \mid x^{(i-1)}\right) \text{ with } \sum_{x^{(i)}} T\left(x^{(i)} \mid x^{(i-1)}\right) = 1 \;\;\; \forall i.
\end{equation}

Possible transitions between the states are specified by this transition matrix.
A Markov chain is said to be homogeneous if its transition probabilities do not change over time, which is indicated by the equation \( T = \Delta \, T\left(x^{(i)} \mid x^{(i-1)}\right) \). An invariant distribution \( p(x) \) refers to a probability distribution that remains unchanged as the Markov chain evolves. This means that regardless of the starting state and after numerous transitions, the chain will eventually settle into this distribution. This behavior follows the objectives when a posterior distribution that cannot be assessed by traditional techniques is approximated using MCMC sampling.
To obtain an invariant distribution, one must build an irreducible, homogeneous, stochastic transition matrix T that is aperiodic. Aperiodicity assures that the chain doesn't get trapped in cycles, while irreducibility ensures that every state can be reached from any other state at some point. Reversibility, often known as the detailed balance requirement, is a sufficient but not necessary condition for the invariance of a target distribution $p(x)$.
\begin{equation}
    p(x^{(i)}) T(x^{(i-1)} \mid x^{(i)}) = p(x^{(i-1)}) T(x^{(i)} \mid x^{(i-1)}).
\end{equation}
Therefore, it is possible to guarantee the invariance of a target distribution p(x) by ensuring a careful balance.

The MCMC analysis is conducted using the \textit{emcee} \citep{Foreman-Mackey:2012any} package. To determine the best-fit parameters, we employ the likelihood function, which is expressed as:
\[
\mathcal{L} \propto \exp\left(-\frac{\chi^2}{2}\right).
\]
It is important to note that minimizing \(\chi^2\) is equivalent to maximizing the likelihood and minimizing the negative log-likelihood.
In this study, we constrain model parameters using the specified data sets. For the MCMC analysis, we compute the \(\chi^2\) function for each data set as follows:
\begin{equation}
    \chi^2_{Data}(\Theta) = \sum_{i=1}^{N} \left[ \frac{(H_{th}(z_i, \Theta) - H_{obs}^{Data}(z_i))^2}{\sigma_H^2(z_i)} \right],
\end{equation}
where \(N\) is the number of data points, \(H_{th}\) represents the theoretical Hubble parameter values for the model with parameters \(\Theta\), \(H_{obs}\) is the observed Hubble parameter from the data, and \(\sigma_H\) denotes the error in the observed Hubble parameter \(H^{Data}\). 

We further analyze the model by combining multiple data sets. To do this, we calculate the total \(\chi^2\) function, which is the sum of the individual \(\chi^2\) functions for each data set:
\begin{equation}
    \chi^2_{Total}(\Theta) = \sum_{i=1}^{N} \chi^2_{D_i}.
\end{equation}
Here, \(D_i\) denotes the different data sets utilized in this analysis.

\section{Results and Discussion}\label{result}
All of the required gear is now in place to advance with our observational analysis of Covariant Extended Proca-Nuevo (CEPN) gravity on a cosmic scale. We carry out the analysis that was mentioned in the preceding section. Further, the constrained model parameters are depicted in \figureautorefname~\ref{fig:combined}, which shows a two-dimensional contour plot at a $99.7\%$ confidence level. These plots indicate that the model aligns well with the observed data. The best-fit values are summarized in \tableautorefname~\ref{tab:tab_1}. For the CEPN model analysis (DESI alone), constrained the sound horizon parameter $r_d$ along with $\Omega_{m_0}$ and $H_0$. This yields the values  
$144.0^{+1.9}_{-2.7}$ for $r_d$ and $0.287 \pm 0.025$ for $\Omega_{m_0}$ depicting the correlation with the variable $H_0$ (see, \figureautorefname~\ref{fig:combined}). However, the sound horizon measured at the drag epoch, which occurs shortly after recombination when photons and baryons decouple, manifests as a distinct peak in the correlation function or as a series of damped oscillations in the power spectrum (for a more detailed discussion, see \citep{BOSS:2016wmc}).

For different datasets, including DESI, DESI+GRBs, DESI+GRBs+CCh, and DESI+GRBs+CCh+SNeIa, the $H(z)$ data and their corresponding error bars are displayed in the first column of \figureautorefname~\ref{fig:cosmographic_parameters1}. The theoretical curve that our CEPN model predicts is represented by the red line in the plot. The close alignment of error bars with these shaded regions indicates a great agreement between the model's predictions and the observed data. The shaded areas indicate different confidence zones (up to 3$\sigma$).

In \figureautorefname~\ref{fig:cosmographic_parameters1} (second column), we focus on the distance modulus function $\mu(z)$. Here, we present an error plot for the observed distance modulus of the 1701 SNeIa and 162 GRBs datasets. The red line depicts the mean theoretical curves obtained from our cosmological model with constrained model parameters by different datasets. The shaded regions represent the error bars at an impressively high confidence level of up to 99.7$\%$. The agreement between the theoretical predictions and the observed distance modulus provides strong support for the accuracy and validation of the CEPN algebraic model in describing the underlying cosmological processes.

Moreover, utilizing the mean values of the constrained parameters, examine the cosmographic parameters like the deceleration parameter $q$. The third column of the \figureautorefname~\ref{fig:cosmographic_parameters1} illustrates the present-day value of $q_{CEPN}$ at $(z=0)$, denoted as $\equiv q_{CEPN,0}$, for the given parameter vector. Thus, the obtained values of $q_{CEPN,0}$ as well as transition redshift $z_t$ are mentioned in the \tableautorefname~\ref{tab:tab_3}. However, this model reveals the quintessence nature of the Universe through this analysis utilizing a data-driven method.

In \figureautorefname~\ref{fig:cosmographic_parameters2}, we also present the $DESI$ data, as discussed in \sectionautorefname~\ref{data} of subsection~\ref{desi}. The figure displays the mean curves (red line) for $D_M$, $D_H$, and $D_V$, each normalized by $r_d$ and plotted against redshift. These curves align perfectly with the visible error bars.

Currently, the $DESI~BAO$ data is highly influential, demonstrating superior constraining power compared to the $Wigglez~BAO$ data (refer \citep{Blake:2011en}). In this study, we utilized this data to constrain the Hubble constant $H_0$, highlighting the dominance of $DESI$. \figureautorefname~\ref{fig:histogram}
depicts the Gaussian and deviation from Gaussian distributions of the corresponding data through histogram analysis. This underscores the significance of the $DESI$ in constraining cosmological model and revealing the nature of the cosmic Universe using the recent data.

\begin{figure}
    \centering
    \includegraphics[width=0.8\linewidth]{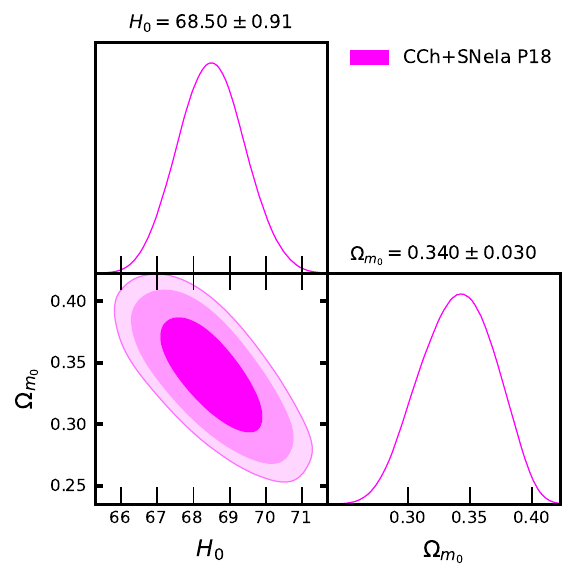}
    \caption{A contour plot showing the model parameters $H_0$ and $\Omega_{m_0}$ obtained through $\chi^2$ analysis for the current model. This plot illustrates the results of a combined analysis of CCh and SNeIa P18 data that consists of 1048 data points.}
    \label{fig:old sneia}
\end{figure}

For the direct comparison of the result with \citep{Anagnostopoulos:2023pvi} we considered the combination of CCh and SNeIa data sets and obtained the results as shown in the \figureautorefname~\ref{fig:old sneia}. It is clearly evident from the values that the scenario presented in \citep{Anagnostopoulos:2023pvi} can be regained when CCh+SNeIa is taken into account.

\begin{table}
    \caption{Results of MCMC for parameters $\Omega_{m_0}$, and $H_0$  (km/s/Mpc)  with $1\sigma$ errors}
    \label{tab:tab_1}
    \centering
    \begin{tabular}{c|c|c}
        \hline
        \hline
        \textbf{Data} &  $\Omega_{m_0}$ & $H_0$  \\ 
        \hline
         DESI& $0.298^{+0.034}_{-0.20}$ & $72.0^{+1.50}_{-0.93}$ \\
          \hline
        DESI+GRBs  & $0.344^{+0.030}_{-0.027}$ &$71.1^{+1.1}_{-1.2}$\\
        \hline
         DESI+GRBs+CCh  & $0.336^{+0.026}_{-0.026}$ & $71.3^{+1.1}_{-1.1}$  \\
          \hline
         DESI+GRBs+CCh+SNeIa  & $0.317^{+0.011}_{-0.012}$ & $73.89^{+0.19}_{-0.19}$  \\
            \hline
            \hline
    \end{tabular}
\end{table}

\begin{table}
    \caption{Present day value ($z=0$) to both deceleration parameter and the transitional redshift}
    \label{tab:tab_3}
    \centering
    \begin{tabular}{c|c|c}
        \hline
        \hline
        \textbf{Data} &  $q_{CEPN,0} $ & $z_t$  \\ 
        \hline
        DESI &  $-0.6364$ & $0.7014$  \\
        \hline
        DESI+GRBs &  $-0.6112$ & $0.6573$  \\
        \hline
        
        DESI+GRBs+CCh & $-0.5858$ & $0.6132$  \\
        \hline
        
        DESI+GRBs+CCh+SNeIa &  $-0.5750$& $0.5971$  \\
        
        \hline
        \hline
    \end{tabular}
\end{table}

\subsection{Analysing Information Criteria} 
As a final step, we utilize the Akaike Information Criterion (AIC) \citep{Hakaike1974}, Bayesian Information Criterion (BIC) \citep{Schwarz:1978tpv}, and The Deviance Information Criterion (DIC) \citep{Spiegelhalter:2002yvw} to compare the efficiency of the model with respect to $\Lambda$CDM and evaluate the compatibility of the observational scenarios.

The AIC is defined as
\begin{equation}
    \text{AIC} = 2k - 2 \ln \mathcal{L}_{\text{max}}.
\end{equation}
Here $\mathcal{L}_{\text{max}}$ represents the maximum likelihood that the model can achieve, and $k$ denotes the number of parameters within the model. The optimal model is the one that minimizes the AIC. The AIC is formulated through an approximate minimization of the Kullback–Leibler information entropy, which quantifies the disparity between the actual data distribution and the distribution predicted by the model. On the other hand, the BIC criterion serves as an estimator of Bayesian evidence, expressed as
\begin{equation}
  \text{BIC} \equiv -2 \ln \mathcal{L}_{\text{max}} + k \ln N, 
\end{equation}
where $N$ represents the number of data points utilized in the fitting process. The BIC is derived from approximating the evidence ratios of models, commonly referred to as the Bayes factor \citep{Kass:1995loi}.

Finally, the DIC criterion is grounded in principles from both Bayesian statistics and information theory \citep{Liddle:2007fy}, and it can be expressed as 
\[
\text{DIC} = {D(\overline \Theta)} + 2\mathcal{P}.
\]
The variable $\mathcal{P}$
 represents the Bayesian complexity, defined as
 \begin{equation}
     \mathcal{P} =\overline {D(\Theta)}-D(\overline{\Theta}),
\end{equation}
where the overline indicates the standard mean value. Additionally, $D(\Theta)$ refers to the Bayesian deviation, a measure closely associated with the effective degrees of freedom. This quantity is expressed as
\begin{equation}
    D(\Theta) = -2 \ln(\mathcal{L}(\Theta)).
\end{equation}
To rank competing models based on their fit to observational data, we focus on the differences in Information Criterion (IC) values. Specifically, we calculate the difference $\Delta IC_{\text{model}} = IC_{\text{model}} - IC_{\text{min}}$, comparing each model's IC to the minimum in the set. According to Jeffrey's scale \citep{Kass:1995loi,Jeffreys:1998h}, if $\Delta IC\leq 2$, the model is statistically compatible with the best one. If $2< \Delta IC< 6$, there is moderate tension, and $\Delta IC \geq 10$ implies a strong tension.

From the \tableautorefname~\ref{tab:ic}, it is evident that our model is consistent with $\Lambda$CDM model which aligns well with observations. All model selection criteria (AIC, BIC, and DIC) support this condition, with the difference of $\Delta IC\leq 2$ in all cases, except for the combined dataset of  DESI+GRBs+CCh+SNeIa. This indicates that the CEPN model is statistically well-compatible with the data in all instances under consideration. However, for the combined DESI+GRBs+CCh+SNeIa dataset, the model exhibits moderate tensions with the $\Lambda$CDM model.

 \begin{table*}
    \caption{Comparative statistical analysis of our model and  $\Lambda$CDM (as reference model) using multiple data sets.}
    \label{tab:ic}
     \setstretch{1.1}
	\centering
	\begin{tabularx}{\linewidth}{>{\centering\arraybackslash}X>{\centering\arraybackslash}X>{\centering\arraybackslash}X >{\centering\arraybackslash}X  >{\centering\arraybackslash}X >{\centering\arraybackslash}X  >{\centering\arraybackslash}X>{\centering\arraybackslash}X>{\centering\arraybackslash}X}
        \hline
        \hline
		\multicolumn{2}{c}{Models} &{$\chi^2_{\text{min}}$} & $AIC$ & $\Delta AIC$ & $BIC$ & $\Delta BIC$ &$ DIC$ &$\Delta DIC$\\ 
		\hline
            \multicolumn{2}{c}{\textbf{ $\Lambda$CDM-model}} \\
            \multicolumn{2}{c}{DESI} & $6.975$ & $10.9754$ & $0.0$ & $13.534$ & $0.0$ & 7.434 & $0.0$\\
            
            \multicolumn{2}{c}{DESI+GRBs} & $233.587$ & $237.587$ & $0.0$  & $239.948$ & $0.0$ & 237.000& $0.0$\\
             
           \multicolumn{2}{c}{DESI+GRBs+CCh} & $248.628$ & $252.628$  & $0.0$ & $255.082$ & $0.0$ & $249.117$& $0.0$\\
            
           \multicolumn{2}{c}{DESI+GRBs+CCh+SNeIa}& $1795.410$ & $1799.41$ & $0.0$ & $1801.969$ & $0.0$ & 1795.509 &$0.0$\\
            \hline
            \multicolumn{2}{c}{\textbf{ CEPN-model}} \\
           \multicolumn{2}{c}{DESI} & $7.3864$ & $11.386$ & $0.4106$ & $13.945$ & $0.411$ & 8.052 & 0.618 \\
            
            \multicolumn{2}{c}{DESI+GRBs} & $233.389$ & $237.389$ & $0.197$  & $240.146$ & $0.198$ & 234.225&  2.775\\
             
           \multicolumn{2}{c}{DESI+GRBs+CCh} & $248.523$ & $252.523$  & $0.105$ & $255.187$ & $0.105$ & 249.050  &  0.067\\
            
           \multicolumn{2}{c}{DESI+GRBs+CCh+SNeIa} & $1799.331$ & $1803.331$ & $3.921$  & $1805.890$ & $3.921$ & 1799.475 & 3.966\\
           \hline
           \hline
	\end{tabularx}
    \end{table*}

\section{Conclusion}\label{conclusion}
In this paper, we addressed cosmological observations of massive Proca-Nuevo gravity.
The first is a recently proposed non-linear theory involving a huge spin-1 field inspired by dRGT massive gravity. It can be extended by including Generalised Proca class operators without breaking the fundamental primary constraint necessary for consistency. After that, the theory can be covariantized and coupled to gravity in a way that produces cosmological solutions that are reliable and ghost-free. Furthermore, compared to usual massive gravity scenarios, these cosmological solutions perform nicely at the perturbative level, showing no signs of instabilities. In a cosmological context, massive Proca-Nuevo gravity adds additional terms to the Friedmann equations, which can be combined into an effective dark-energy component.

In this study, we have obtained the constraints of the free parameters of the theory by a model-independent and data-driven technique employing DESI, CCh, GRBs, and SNeIa data. Specifically, we are provided with the parameters' corresponding likelihood contours, error bar plots, best-fit values, and a confidence level of up to 3$\sigma$, indicating that the approach agrees with observations. Eventually, statistical consistency with the data simulation was demonstrated by analyzing the deceleration parameter at its current value. The results indicate that the Universe's expansion is quintessence, as reflected in the obtained values of $q_{EPN,0}$. Furthermore, the efficiency and compatibility of the CEPN model with observational data, compared to the $\Lambda$CDM model, are confirmed using the information criteria method.

As per our observation, using DESI BAO dataset constrains the value of the Hubble constant comparatively higher than other BAO surveys. This probably can lead to new milestones in alleviating $H_0$ tension. Additionally, as we have presented, its power of constraining is also good. On comparing the BAO data from WiggleZ and DESI surveys, we have observed the Gaussianity in the distribution and corresponding constraining ability on the Hubble constant of the associated DESI data.  Further, the value of $H_0$ is a bit on the higher side, particularly for one combination with the value being $73.89 km/s/Mpc$. This is obtained for the combined dataset of DESI, GRBs, CCh, and Pantheon+SH0ES. The higher drag is mainly because of the presence of Cepheid measurements along with DESI. However, we believe that DESI results act significant in alleviating tension. Some interesting literature in this regard can be seen in \citep{Banerjee:2022ynv, Ren:2022aeo, Pan:2019gop, Petronikolou:2021shp, Petronikolou:2023cwu, Basilakos:2023kvk, Saridakis:2021xqy}.

Analyzing the interactions at a perturbative level, utilizing data from Large-scale Structures (such as f$\sigma_8$ measurements) and additional methods, is beyond the scope of this current study. This will be addressed in future research.

\section*{Acknowledgements}
LS, NSK, and VV acknowledge DST, New Delhi, India, for supporting research facilities under DST-FIST-2019. LS  acknowledges Kuvempu University for providing University General Fellowship (File no. KU: B.C.M-3/145/2023-24 Dated: 04/08/2023). We would like to express our gratitude to the anonymous referee for providing thoughtful comments and suggestions on our manuscript.
 
\section*{Data Availability}

There are no new data associated with this article.

\bibliographystyle{apj.bst}
\bibliography{apj} 


\end{document}